\documentclass[12pt]{article}
\usepackage{epsfig}

\usepackage{rotating}

\usepackage{epsfig}
\usepackage{bm}

\title{$Q^2$-evolution of parton densities at small $x$
values.\\
Combined H1$\&$ZEUS F2 data. 
\footnote{The work was supported by RFBR grant No.11-02-1454-a}}
\author{  A.V. Kotikov and B.G. Shaikhatdenov\\
Joint Institute for Nuclear Research, Russia \\
}

\begin{document}
\maketitle

\abstract{
We use the Bessel-inspired behavior of
the structure function $F_2$ at small $x$,
obtained for a flat initial condition in the DGLAP evolution equations,
with ``frozen'' and analytic modifications of the
strong coupling constant to study precise
combined H1$\&$ZEUS data for the structure function $F_2$
published recently.
}



\section{Introduction}
A reasonable agreement between HERA data  \cite{H1ZEUS}-\cite{DIS02}
and the next-to-leading-order (NLO) approximation of
perturbative Quantum Chromodynamics (QCD)
has been observed for $Q^2 \geq 2$ GeV$^2$ (see reviews in \cite{CoDeRo}
and references therein), which gives us a reason to believe that
perturbative QCD is capable of describing the
evolution of the structure function (SF) $F_2$ and its derivatives
down to very low $Q^2$ values, where all the strong interactions
are conventionally considered to be soft processes.

A standard way to study the $x$ behavior of
quarks and gluons is to compare the data
with the numerical solution to the
Dokshitzer-Gribov-Lipatov-Altarelli-Parisi (DGLAP)
equations
\cite{DGLAP}
by fitting the parameters of
$x$-profile of partons at some initial $Q_0^2$ and
the QCD energy scale $\Lambda$ \cite{fits,Ourfits}.
However, for the purpose of analyzing exclusively the
small-$x$ region, there is the alternative of doing a simpler analysis
by using some of the existing analytical solutions to DGLAP equations
in the small-$x$ limit \cite{BF1,Q2evo}.

The ZEUS and H1 Collaborations have presented
the new precise combined data \cite{Aaron:2009aa} for the SF
$F_2$.
The aim of this short paper is to compare
the combined H1$\&$ZEUS data with
the predictions obtained by using the so-called doubled asymptotic scaling (DAS)
approach \cite{Q2evo}.

To improve the analysis at low $Q^2$ values, it is important to consider
the well-known infrared modifications of the strong coupling
constant. We will use the ``frozen'' and analytic versions (see,
\cite{Cvetic1} and references therein).

\section{Parton distributions and the structure function $F_2$
}

Here, for simplicity we consider only  the
leading order (LO)
approximation\footnote{
The NLO results can be found in  \cite{Q2evo}.}.
The structure function $F_2$ has the form
\begin{eqnarray}
    F_2(x,Q^2) &=& e \, f_q(x,Q^2),~~
    f_a(x,Q^2)
~=~ f_a^{+}(x,Q^2) + f_a^{-}(x,Q^2),~~(a=q,g)
\label{8a}
\end{eqnarray}
where
$e=(\sum_1^f e_i^2)/f$ is an average charge squared.

The small-$x$ asymptotic expressions for parton densities
$f^{\pm}_a$ look like
\begin{eqnarray}
    f^{+}_g(x,Q^2) &=& \biggl(A_g + \frac{4}{9} A_q \biggl)
        I_0(\sigma) \; e^{-\overline d_{+} s} + O(\rho),~~
    f^{+}_q(x,Q^2) ~=~
\frac{f}{9} \frac{\rho I_1(\sigma)}{I_0(\sigma)}
+ O(\rho),
\nonumber \\
    f^{-}_g(x,Q^2) &=& -\frac{4}{9} A_q e^{- d_{-} s} \, + \, O(x),~~
    f^{-}_q(x,Q^2)
~=~ A_q e^{-d_{-}(1) s} \, + \, O(x),
    \label{8.02}
\end{eqnarray}
where $I_{\nu}$ ($\nu=0,1$)  are the modified Bessel
functions,
\begin{equation}
s=\ln \left( \frac{a_s(Q^2_0)}{a_s(Q^2)} \right),~~
\sigma = 2\sqrt{\left|\hat{d}_+\right| s
 \ln \left( \frac{1}{x} \right)}  \ ,~~~ \rho=\frac{\sigma}{2\ln(1/x)} \ ,
\label{intro:1a}
\end{equation}
and
\begin{equation}
\hat{d}_+ = - \frac{12}{\beta_0},~~~
\overline d_{+} = 1 + \frac{20f}{27\beta_0},~~~
d_{-} = \frac{16f}{27\beta_0}
\label{intro:1b}
\end{equation}
denote singular and regular parts of the anomalous dimensions
$d_{+}(n)$ and $d_{-}(n)$,
respectively, in the limit $n\to1$\footnote{
We denote the singular and regular parts of a given quantity $k(n)$ in the
limit $n\to1$ by $\hat k/(n-1)$ and $\overline k$, respectively.}.
Here $n$ is a variable in the Mellin space.

\begin{figure}[t]
\centering
\vskip 0.5cm
\includegraphics[height=0.75\textheight,width=1.05\hsize]{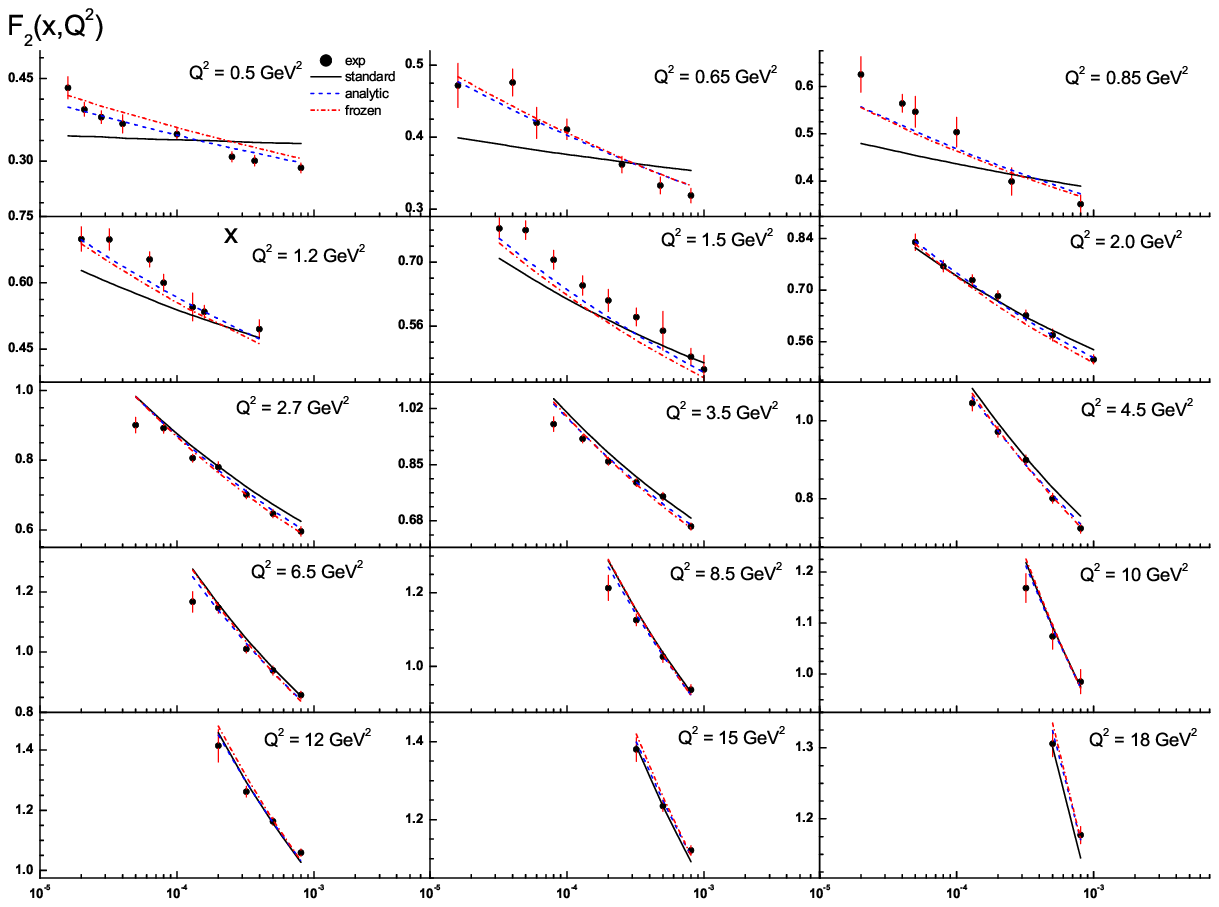}
\vskip -0.3cm
\caption{$x$ dependence of $F_2(x,Q^2)$ in bins of $Q^2$.
The combined experimental data from H1 and ZEUS Collaborations
\cite{Aaron:2009aa} are
compared with the NLO fits for $Q^2\geq0.5$~GeV$^2$ implemented with the
standard (solid lines), frozen (dot-dashed lines), and analytic (dashed lines)
versions of the strong coupling constant.}
\label{fig:F1}
\end{figure}

\section{``Frozen'' and analytic  coupling constants}

In order to improve an agreement at low $Q^2$ values,
the QCD coupling constant is modified in the infrared region.
We consider two modifications that effectively increase the argument of the coupling constant
at low $Q^2$ values (see \cite{DoShi}).

In the first case, which is more phenomenological, we introduce freezing
of the coupling constant by changing its argument $Q^2 \to Q^2 + M^2_{\rho}$,
where $M_{\rho}$ is the $\rho $-meson mass (see \cite{Cvetic1} and
discussions therein). Thus, in the
formulae of Sec. 2 we have to carry out the following replacement:
\begin{equation}
 a_s(Q^2) \to a_{\rm fr}(Q^2) \equiv a_s(Q^2 + M^2_{\rho})
\label{Intro:2}
\end{equation}

The second possibility follows the Shirkov--Solovtsov idea
\cite{ShiSo}
concerning the analyticity of the coupling constant that leads
to additional power dependence of the latter.
Then, in the formulae of the previous section
the coupling constant $a_s(Q^2)$ should be replaced as follows:
\begin{eqnarray}
 a^{\rm LO}_{\rm an}(Q^2) \, = \, a_s(Q^2) - \frac{1}{\beta_0}
 \frac{\Lambda^2_{\rm LO}}{Q^2 - \Lambda^2_{\rm LO}}
\label{an:LO}
\end{eqnarray}
in the LO approximation and
\begin{eqnarray}
 a_{\rm an}(Q^2) \, = \, a_s(Q^2) - \frac{1}{2\beta_0}
 \frac{\Lambda^2}{Q^2 - \Lambda^2}
+ \ldots \, ,
\label{an:NLO}
\end{eqnarray}
in the NLO approximation. Here the the symbol $\ldots$ stands
for the terms that provide negligible contributions when $Q^2 \geq 1$ GeV \cite{ShiSo}.

Note here that the perturbative coupling constant $a_s(Q^2)$
is different in the LO
and NLO approximations. Indeed, from the renormalization group equation
we can obtain the following equations for the coupling constant
\begin{eqnarray}
 \frac{1}{a_s^{\rm LO}(Q^2)} \, = \, \beta_0
 \ln{\left(\frac{Q^2}{\Lambda^2_{\rm LO}}\right)}
\label{as:LO}
\end{eqnarray}
in the LO approximation and
\begin{eqnarray}
 \frac{1}{a_s(Q^2)} \, + \,
 \frac{\beta_1}{\beta_0} \ln{\left[
 \frac{\beta_0^2 a_s(Q^2)}{\beta_0+ \beta_1 a_s(Q^2)}\right]} \, = \,
 \beta_0 \ln{\left(\frac{Q^2}{\Lambda^2}\right)}
\label{as:NLO}
\end{eqnarray}
in the NLO approximation.
Usually at the NLO level ${\rm \overline{MS}}$-scheme is used;
therefore, below we apply $\Lambda = \Lambda_{\rm \overline{MS}}$.

\section{Comparison with experimental data
} \indent

By using the results of the previous section we have
analyzed H1$\&$ZEUS data for $F_2$ \cite{Aaron:2009aa}.
%
In order to keep the analysis as simple as possible,
we fix $f=4$ and $\alpha_s(M^2_Z)=0.1168$ (i.e., $\Lambda^{(4)} = 284$ MeV) in agreement
with more recent ZEUS results given in~\cite{H1ZEUS}.

\begin{table}
\caption{
The results of LO and NLO fits to  H1 $\&$ ZEUS data
\cite{Aaron:2009aa},
with various lower cuts on $Q^2$; in the fits
the number of flavors $f$ is fixed to 4.
}
\centering
\footnotesize
\vspace{0.3cm}
\begin{tabular}{|l||c|c|c||r|} \hline \hline
& $A_g$ & $A_q$ & $Q_0^2~[{\rm GeV}^2]$ &
 $\chi^2 / n.d.f.$~ \\
\hline\hline
~$Q^2 \geq 5 {\rm GeV}^2 $  &&&& \\
 LO & 0.623$\pm$0.055 & 1.204$\pm$0.093 & 0.437$\pm$0.022 & 1.00 \\
 LO$\&$an. & 0.796$\pm$0.059 & 1.103$\pm$0.095 & 0.494$\pm$0.024 & 0.85  \\
  LO$\&$fr. & 0.782$\pm$0.058 & 1.110$\pm$0.094 & 0.485$\pm$0.024 & 0.82   \\
\hline
 NLO & -0.252$\pm$0.041 & 1.335$\pm$0.100 & 0.700$\pm$0.044 & 1.05 \\
 NLO$\&$an. & 0.102$\pm$0.046 & 1.029$\pm$0.106 & 1.017$\pm$0.060 & 0.74  \\
  NLO$\&$fr. & -0.132$\pm$0.043 & 1.219$\pm$0.102 & 0.793$\pm$0.049 & 0.86   \\
\hline\hline
~$Q^2 \geq 3.5 {\rm GeV}^2 $  &&&& \\
 LO & 0.542$\pm$0.028 & 1.089$\pm$0.055 & 0.369$\pm$0.011 & 1.73 \\
 LO$\&$an. & 0.758$\pm$0.031 & 0.962$\pm$0.056 & 0.433$\pm$0.013 & 1.32  \\
  LO$\&$fr. & 0.775$\pm$0.031 & 0.950$\pm$0.056 & 0.432$\pm$0.013 & 1.23   \\
\hline
 NLO & -0.310$\pm$0.021 & 1.246$\pm$0.058 & 0.556$\pm$0.023 & 1.82 \\
 NLO$\&$an. & 0.116$\pm$0.024 & 0.867$\pm$0.064 & 0.909$\pm$0.330 & 1.04  \\
  NLO$\&$fr. & -0.135$\pm$0.022 & 1.067$\pm$0.061 & 0.678$\pm$0.026 & 1.27 \\
\hline \hline
~$Q^2 \geq 2.5 {\rm GeV}^2 $  &&&& \\
 LO & 0.526$\pm$0.023 & 1.049$\pm$0.045 & 0.352$\pm$0.009 & 1.87 \\
 LO$\&$an. & 0.761$\pm$0.025 & 0.919$\pm$0.046 & 0.422$\pm$0.010 & 1.38  \\
  LO$\&$fr. & 0.794$\pm$0.025 & 0.900$\pm$0.047 & 0.425$\pm$0.010 & 1.30   \\
\hline
 NLO & -0.322$\pm$0.017 & 1.212$\pm$0.048 & 0.517$\pm$0.018 & 2.00 \\
 NLO$\&$an. & 0.132$\pm$0.020 & 0.825$\pm$0.053 & 0.898$\pm$0.026 & 1.09  \\
  NLO$\&$fr. & -0.123$\pm$0.018 & 1.016$\pm$0.051 & 0.658$\pm$0.021 & 1.31   \\
\hline\hline
~$Q^2 \geq 0.5 {\rm GeV}^2 $  &&&& \\
 LO & 0.366$\pm$0.011 & 1.052$\pm$0.016 & 0.295$\pm$0.005 & 5.74 \\
 LO$\&$an. & 0.665$\pm$0.012 & 0.804$\pm$0.019 & 0.356$\pm$0.006 & 3.13  \\
  LO$\&$fr. & 0.874$\pm$0.012 & 0.575$\pm$0.021 & 0.368$\pm$0.006 & 2.96   \\
\hline
 NLO & -0.443$\pm$0.008 & 1.260$\pm$0.012 & 0.387$\pm$0.010 & 6.62 \\
 NLO$\&$an. & 0.121$\pm$0.008 & 0.656$\pm$0.024 & 0.764$\pm$0.015 & 1.84  \\
  NLO$\&$fr. & -0.071$\pm$0.007 & 0.712$\pm$0.023 & 0.529$\pm$0.011 & 2.79 \\
\hline \hline
\end{tabular}
\end{table}

As can be seen from Fig.~1 and Table~1,
the twist-two approximation is reasonable for $Q^2 \geq 4$ GeV$^2$.
At lower $Q^2$ we observe that the fits in the cases with ``frozen'' and
analytic strong coupling constants are very similar
(see also \cite{KoLiZo,Cvetic1}) and describe the data in
the low $Q^2$ region significantly better than the standard fit.
Nevertheless, for $Q^2 \leq 1.5$~GeV$^2$
there is still some disagreement with
the data, which needs to be additionally studied.
In particular,  the Balitsky--Fadin--Kuraev--Lipatov (BFKL)
resummation \cite{BFKL} may be important here \cite{Kowalski:2012ur}.
It can be added in the generalized DAS approach according to the discussion
in Ref. \cite{KoBaldin}.

\section{Conclusions} \indent
We have studied
the $Q^2$-dependence of the structure function $F_2$
at small-$x$ values within the
framework of perturbative QCD. Our twist-two
results are well consistent with
precise H1$\&$ZEUS data
\cite{Aaron:2009aa}
in the region of $Q^2 \geq 4$~GeV$^2$,
where perturbative theory is thought to be applicable.
The usage of ``frozen'' and analytic modifications of the
strong coupling constant,
$\alpha_{\rm fr}(Q^2)$ and $\alpha_{\rm an}(Q^2)$,
is seen to improve an agreement with experiment at low $Q^2$ values,
$Q^2 \leq 1.5$~GeV$^2$.\\




\end{document}